# Minimizing Maintenance Cost Involving Flow-time and Tardiness Penalty with Unequal Release Dates

Kondo Hloindo Adjallah[(1)*] and Kossi Pelope Adzakpa[(2)**]

[(1)]Institut Charles Delaunay, University of Technology of Troyes
12 rue Marie Curie, 10010 Troyes cedex, France
[(2)]Hydrogen Research Institute, Université du Québec à Trois-Rivières, CP 500
Trois-Rivères, (Québec), G9A 5H7, Canada

**Abstract**

This paper proposes important and useful results relating to the minimization of the sum of the flow-time and the tardiness of tasks or jobs with unequal release dates (occurrence date), with application to maintenance planning and scheduling. Firstly, the policy of real-time maintenance is defined for minimizing the cost of tardiness and critical states. The required local optimality rule (FTR) is proved, in order to minimize the sum or the linear combination of the tasks' flow-time and tardiness costs. This rule has served to design a scheduling algorithm, with $O(n^3)$ complexity when it is applied to schedule a set of $n$ tasks on one processor. To evaluate its performance, the results are compared to a lower bound that is provided, in a numerical case study. Using this algorithm in combination with the tasks urgency criterion, a real-time algorithm is developed to schedule the tasks on $q$ parallel processors. This latter algorithm is finally applied to schedule and assign preventive maintenance tasks to processors in the case of a distributed system. Its efficiency enables, as shown in the numerical example, to minimize the cost of preventive maintenance tasks, expressed as the sum of the tasks tardiness and flow-time. This corresponds to costs of critical states and of tardiness of preventive maintenance.

**Notations**

| | | | |
|---|---|---|---|
| $N$ | number of sites in the system | $\alpha_{ik}^1$ | first threshold on machine $E_{ik}$ |
| $N_k$ | number of machines on site $k$ | $\alpha_{ik}^2$ | second threshold on machine $E_{ik}$ |
| $E_{ik}$ | machine $i$ of site $k$ | $\tau_{ik}^1$ | duration related to threshold $\alpha_{ik}^1$ |
| $A_{ik}(t)$ | availability of machine $E_{ik}$ at time $t$ | $\tau_{ik}^2$ | duration related to threshold $\alpha_{ik}^2$ |
| $H$ | analysis period | $\lambda_{ik}$ | failure rate of machine $E_{ik}$ |
| $C_{tot}$ | total cost on period $H$ | $\mu_{ik}$ | repair rate of machine $E_{ik}$ |
| $r_{m,ik}$ | tasks' release dates on machine $E_{ik}$ | $\beta_{ik}$ | Weibull parameter related to machine $E_{ik}$ |
| $d_{m,ik}$ | tasks' due dates on machine $E_{ik}$ | $\gamma_{ik}$ | Weibull parameter related to machine $E_{ik}$ |
| $c_{m,ik}$ | tasks' completion dates on machine $E_{ik}$ | $\sigma_{ik}$ | Weibull parameter related to machine $E_{ik}$ |
| $f_i$ | tasks' strength vector | $\Omega(i, j)$ | tasks' dominance matrix |

## 1. INTRODUCTION

The modelling and optimization of maintenance process of interest to academic researchers as well as industry managers. They seek optimal management solutions enabling the machines and devices to reach the highest availability and efficiency, which are necessary conditions to provide high quality products and services to customers. The increasing number of publications in this field during the last decade is a witness of this interest. In the literature, most of maintenance process models are stochastic models often based on state graphs. Most of them relate to the study of systems comprising not more than three components. As usual, the goals are to minimize economic objectives on a time horizon that is sometimes supposed to be boundless. Because of the specificity of the problem and of the various models used to describe the maintenance process, various resolution methods are used.

---

[*]Corresponding author: *kondo.adjallah@utt.fr*
[**]This work was done when (2) was at the University of Technology of Troyes.

Whereas works of Calibra *et al.* [1] and Fyffe *et al.* [2] focus on the problem of reliability allocation, in design phase, to components of machines to reduce the maintenance cost, works of Berg [3], Park [4] and Stadje and Zuckerman [5] aim at reducing the same cost, through reliability and availability improvement and through optimal replacement of components during exploitation of the machine. Working for the same objectives, Lam [6] proposed a solution based on optimal combination of old and new components. Whilst the maintenance cost optimization can also be achieved through optimal resources allocation to tasks, the above mentioned approaches do not consider maintenance job management problems. On the basis of queuing theory and stochastic models, Stadje and Zuckerman [5] proposed a solution of repairmen allocation to failed machines. A solution based on the queuing theory and a Markov model was also proposed by Derman [7], Frostig [8], [9], [10], Koole [11] and Seshadri [12].

Among the works considering the specific problem of maintenance tasks scheduling and resources allocation, one should consider the works of Graves and Lee [13], Lee and Chen [14], Qi *et al.* [15] and Weinstein and Chung [16]. Their approaches belong to the operations research area. The common aspect of these works is the integration of maintenance policies in production management. Adzakpa *et al.* [17], [18] also proposed in this framework an approach for real-time scheduling and assignment of tasks to processors in a preventive maintenance policy. These algorithms use local optimality rules for the flow-time minimization, and are adapted to the NP-hard nature of the problem – it cannot be solved within a deterministic polynomial time, using an optimal algorithm at the sense of the complexity theory – and to the problem of real-time preventive maintenance decision making.

It is well known that the management cost of the maintenance actions depends on the components' reliability and maintainability characteristics, defined in the design phase, on the number of allocated maintenance task processors and their skills, as well as on the adopted policy and maintenance strategies. The preventive maintenance cost is related to costs of the critical states and to tardiness of machines preventive maintenance activities. Most of maintenance models in the literature propose optimal periods for the preventive maintenance tasks, in order to comply with given availability objectives at a minimum cost. In general, these methods make very strong assumptions about the resources or processors availability for the maintenance tasks: resources available as soon as they are needed, instantly repaired machines and so on. Very often, the assumptions are too hard for the realistic conditions in industry. Obviously, it is more difficult in practice to face the failures of a whole system of reparable machines with a limited number of maintenance tasks processors. Indeed, a task processor is not always available at the planned date. Moreover, incidents may delay the planned maintenance actions and contribute to degrading the machines and thus increase the maintenance costs. Compared to the state of the art, this paper proposes a solution to realistic cases of system comprising *N* machines, where the joint problems of scheduling and real-time assignment to processors of the preventive maintenance tasks are considered. In this case, comparatively to production planning, the maintenance cost minimization implies the flow-time cost minimization with unequal release dates (tasks occurrence dates) under tardiness penalty, while assuming that the processing cost on each machine is constant on the considered time horizon. Parts of these problems have been considered in the production scheduling area ([19], [20], [21], [22]).

This paper presents an algorithm for real-time maintenance tasks planning in distributed systems. It enables to minimize the preventive maintenance cost, expressed as the sum of the tardiness and flow-time costs. The proposed algorithm is based on a priority rule locally optimal with proof. Section 2 presents the real-time maintenance problem considered in this paper. Section 3 deals with the priority rule in a static scheduling process, a lower bound for performance evaluation and an algorithm of the management decision making. In section 4, the priority rule is applied to the problem of real-time planning of maintenance tasks with a real-time scheduling algorithm, emphasized by experimental numerical results. Lastly, section 5 concludes the paper with some indications on the future extensions.

## 2. THE REAL-TIME MAINTENANCE PROBLEM

The aim of this work is to schedule and assign tasks to processors while minimizing the total maintenance actions cost on a finite horizon $H$, under availability constraint.

### 2.1. Formulation of the problem

One considers a distributed system composed of $N$ sites working in parallel and sharing $q$ processors for the preventive maintenance activities of all the machines of the system. On a given exploitation site, the machines operate in series. A machine may be a production machine or just a component. Obviously, the number of processors is less than the total number of machines in the system $\left(q << \sum_{k=1}^{N} N_k\right)$ and the processors are shared by the whole system, $N_k$ being the number of machines on site $k$. The set-up times of tasks (task preparation times) are assumed to be included in their processing time.

To model this problem, finite time horizons are considered, and unavailability and maintenance cost, pertaining to each site must be minimized within the finite horizon. The reliability models of machines in the system are assumed to be known. Perfect maintenance is assumed, i.e., at the end of each preventive maintenance action at time $T$ the machine $E_{ik}$ is as new, with an availability reset to its initial value. Moreover, for a machine, one assumes the maintenance task completion time is also the time of restart. It should be recalled that, if, for a machine, the failure and repair rates are constant on the horizon $H$, then the reliability and availability of this machine depend on the exponential law. In this context, the availability of a machine $E_{ik}$ (machine $i$ of site $k$) after the completion of a preventive maintenance task at time $T$ is expressed as [23]:

$$A_{ik}(t) = \frac{\mu_{ik}}{\lambda_{ik} + \mu_{ik}} + \frac{\lambda_{ik}}{\lambda_{ik} + \mu_{ik}} \exp\left[-(\lambda_{ik} + \mu_{ik})(t - T)\right] \tag{1}$$

where the failure and repair rates ($\lambda_{ik}$ and $\mu_{ik}$) of $E_{ik}$ are constant. The operational availability limit is then defined for each machine, which is converted into a threshold $\alpha_{ik}^1$ such that $A_{ik} \geq \alpha_{ik}^1$. Solving this inequality yields the duration $\tau_{ik}^1$ between two consecutive maintenance tasks on $E_{ik}$

$$\tau_{ik}^1 = \frac{-1}{\lambda_{ik} + \mu_{ik}} \ln\left[\alpha_{ik}^1\left(1 + \frac{\mu_{ik}}{\lambda_{ik}}\right) - \frac{\mu_{ik}}{\lambda_{ik}}\right] \tag{2}$$

Under this threshold, when a machine's working time exceeds $\tau_{ik}^1$ (i.e. $A_{ik} \leq \alpha_{ik}^1$), one supposes that it operates in a critical state beyond this moment with a high probability of failure. Beyond a second threshold $\alpha_{ik}^2$ ($\alpha_{ik}^2 < \alpha_{ik}^1$), there are additional costs to the maintenance action. These last costs are expressed as tardiness costs. The time from the completion date of a maintenance task on each machine $E_{ik}$ (supposed to be a restarting date of the machine) to the deadline of the next task is then

$$\tau_{ik}^2 = \frac{-1}{\lambda_{ik} + \mu_{ik}} \ln\left[\alpha_{ik}^2\left(1 + \frac{\mu_{ik}}{\lambda_{ik}}\right) - \frac{\mu_{ik}}{\lambda_{ik}}\right] \tag{3}$$

In a more general case where machine $E_{ik}$'s reliability follows the Weibull law, its availability $A_{ik}$ is renewed at the end of a maintenance task completed at time $T$. This availability is then $A_{ik}(t-T)$ where $A_{ik}(t)$ is expressed with the availability function below with a constant repair rate $\mu$ (which is $\mu_{ik}$ for machine $E_{ik}$) [24]

$$A(t) = \exp\left(-\left[\mu.(t-\gamma) + \left(\frac{t-\gamma}{\sigma}\right)^\beta\right]\right) \times \left(1 + \mu\sigma.\int_0^{\frac{t-\gamma}{\sigma}} \exp(\mu\sigma x + x^\beta)dx\right) \tag{4}$$

where the Weibull parameters relative to machine $E_{ik}$ are

- $\gamma_{ik}$ the time origin parameter (denoted $\gamma$ in general)

- $\sigma_{ik} > 0$ the scale parameter (denoted $\sigma$ in general)

- $\beta_{ik} > 0$ the shape parameter (denoted $\beta$ in general).

In this case and when the shape parameter $\beta_{ik}$ is greater than or equal to 1, the availability function is a strictly decreasing function which tends to a limit $A_{ik}(\infty)$ when the time $t$ tends to infinity. Thus it exists a unique duration $\tau_{ik}^1$ (resp. $\tau_{ik}^2$) such that $A_{ik}(\tau_{ik}^1) = \alpha_{ik}^1$ (resp. $A_{ik}(\tau_{ik}^2) = \alpha_{ik}^2$). These durations $\tau_{ik}^1$ and $\tau_{ik}^2$ can be computed from the availability function.

The total cost due to the critical states and to the tardiness of the preventive maintenance tasks on the considered time horizon $H$ is then

$$C_{tot}(H) = \sum_{k=1}^{N}\sum_{i=1}^{N_k}\left(\sum_{m/r_{m,ik} \in H}\left((c_{m,ik} - r_{m,ik}) + \max(0, c_{m,ik} - d_{m,ik})\right)\right) \tag{5}$$

where $r_{m,ik}$, $c_{m,ik}$ and $d_{m,ik}$ are respectively the release date (task occurrence date), the completion date and the due date of the $m^{th}$ preventive maintenance task on machine $E_{ik}$. The processing time of the corresponding task is supposed to be the mean time to preventive maintenance of the machine, which is $1/\mu_{ik}$.

**Remark 1:** In the cost $C_{tot}(H)$, the flow-time cost and the tardiness cost are supposed to be equally weighted. When they are unequally weighted with the weight $W_f$ for the flow-time cost and $W_t$ for the tardiness cost, then $C_{tot}(H)$ is replaced by

$$C_{tot}^w(H) = \sum_{k=1}^{N}\sum_{i=1}^{N_k}\left(\sum_{m/r_{m,ik} \in H} W_f.((c_{m,ik} - r_{m,ik}) + W_t.\max(0, c_{m,ik} - d_{m,ik}))\right) \tag{6}$$

A corresponding remark in the sequel allows to still use the approach in its weighted form.

The problem is then

**(P1):** *Minimize* $C_{tot}(H)$ *with the q processors and* $A_{ik} \geq \alpha_{ik}^1$, $k \in \{1,...,N\}$, $i \in \{1,...,N_k\}$

and in its weighted form,

**(P2):** *Minimize* $C_{tot}^w(H)$ *with the q processors and* $A_{ik} \geq \alpha_{ik}^1$, $k \in \{1,...,N\}$, $i \in \{1,...,N_k\}$.

When there is only the flow-time (response-time) cost or critical state cost, and no tardiness cost, and the machines in the system are equally weighted, the problem was dealt with in [17]. The proofs of the following propositions are also based on the same reference. When the different machines in the system are weighted, the problem was considered in [18]. Problems (P1) and (P2) have the properties in the propositions below. The first one is straightforward. The second one is proved as a consequence of an analogous proposition in [17].

**Proposition 1:** *The tasks relative to machine $E_{ik}$ are linked by precedence relationships defined as follow. If $c_{m,ik}$ is the completion time of the $m^{th}$ preventive maintenance task on machine $E_{ik}$ (its $m^{th}$ restarting date), then the release date (resp. the due date) of the $(m+1)^{th}$ task is determined by $r_{m+1,ik} = c_{m,ik} + \tau_{ik}^1$ (resp. $d_{m+1,ik} = c_{m,ik} + \tau_{ik}^2$) with $\tau_{ik}^1$ defined in equations (2) and (3).*

**Proposition 2:** *Problem (P1) is NP-hard.*

In the solution, the tasks are processed without pre-emption, i.e. once started, each maintenance tasks is completely processed without interruption. It is based on the development in the following section on an off-line problem and subsequent properties for a real-time use.

## 3. THE SCHEDULING PROBLEM AND ITS RESOLUTION

In order to solve the cost minimization problem of real-time preventive maintenance planning and scheduling, one considers, in a first step, the static problem in this section. To this end, one seeks to schedule a set of *n* tasks having unequal release dates on a single processor or on parallel processors, in order to minimize, through a non-preemtive schedule, an objective function which is a linear combination of the total flow-time and the total tardiness: $\sum_{i=1}^{n} W_f .(c_i - r_i) + W_t . \max(0, c_i - d_i)$, where $c_i$ is the completion time of task *i*. Each task *i* has a release date $r_i$ (the time the task is released for processing), a processing time $p_i$, and a due date $d_i$; beyond this last date, the task involves a tardiness penalty. $W_f$ and $W_t$ are respectively the weight of the critical states and the weight of the tardiness.

The problem is NP-hard, even in the case considered in this paper where $W_f=W_t=1$ is assumed. In order to solve it, some tools are provided for local optimality and used with the concepts introduced in the following subsection.

### 3.1. Concepts and definitions

**Definition 1:** *Consider a pair of tasks $\{i, j\}$ to be scheduled at time t at the end of a partial sequence $[\pi]$.*

i) It is said that a task *i* *dominates* a task *j* at time *t* and noted $i \prec j$, if scheduling *i* at the end of $[\pi]$ before *j*, yields a less great value of the total cost (here the total weighted sum of the flow-time and the tardiness) than scheduling *j* before *i*.

ii) A matrix denoted $\Omega$ related to the tasks *i* and *j* is then created as follow:

$$\Omega(i,j) = \begin{cases} 1 & \text{if } i \neq j \text{ and } i \prec j \\ 0 & \text{otherwise} \end{cases} \quad (7)$$

**Remark 2:** In the classical scheduling vocabulary, the term dominance is used for partial schedules. But in the above definition of dominance, it can be noticed that if at time *t*, $[\pi]$ is a partial schedule, saying $i \prec j$ is equivalent to saying that the sequence $[\pi ji]$ dominates $[\pi ij]$.

**Definition 2:** *At any time t, the strength $f_i$ of a task i is defined as the number of tasks that i dominates. In other words,*

$$f_i = \sum_j \Omega(i,j) \qquad (8)$$

Now one considers a pair of tasks $\{i, j\}$ to be scheduled at time $t$ at the end of a partial schedule $[\pi]$.

**Remark 3:** Ties are broken with the tasks' indices. So, if $i \neq j$, one does not have $i \prec j$ and $j \prec i$.

**Definition 3:** At any time $t$, the function *FTR* (Flow-time and Tardiness Rule) relatively to a pair of tasks $\{i, j\}$ is defined as

$$FTR(i, j, t) = max[PRTF(i, t), Q_1(i, j, t)] + max[PRTT(i, t), Q_2(i, j, t)] \qquad (9)$$

where

$PRTF(i, t) = 2.max(r_i, t) + p_i,$

$PRTT(i, t) = max(r_i, t) + max(max(r_i, t) + p_i, d_i),$

$Q_1(i, j, t) = max(r_i, t) + max(r_j, t)$ and

$Q_2(i, j, t) = max(max(r_i, t), d_i - p_i) + max(max(r_j, t), d_j - p_j).$

Then the theorem below can be proved.

**Theorem 1:** *At any time $t$ at the end of a partial schedule $[\pi]$ and for any pairs of tasks $\{i, j\}$, $i$ dominates $j$ if and only if*

$$FTR(i, j, t) \leq FTR(j, i, t) \qquad (10)$$

*Proof:* Task $i$ is defined by the parameters $r_i$, $p_i$ and $d_i$. Task $j$ is defined by the parameters $r_j$, $p_j$ and $d_j$.

By scheduling $i$ and $j$ at time $t$ at the end of the partial sequence $[\pi]$, let's denote $C_{ij}$ the additional cost involved by scheduling $i$ directly before $j$, and $C_{ji}$ the one obtained by scheduling $j$ directly before $i$. In this case, if one poses $R_i = max(r_i, t)$ and $R_j\ max(r_j, t)$, then

$$\begin{aligned} C_{ij} = & \{[R_i + p_i - r_i] + [max(R_i + p_i - d_i, 0)]\} + \\ & \{[max(R_i + p_i, R_j) + p_j - r_j] + [max(max(R_i + p_i, R_j) + p_j - d_j, 0)]\} \end{aligned}$$

and

$$\begin{aligned} C_{ji} = & \{[R_j + p_j - r_j] + [max(R_j + p_j - d_j, 0)]\} + \\ & \{[max(R_j + p_j, R_i) + p_i - r_i] + [max(max(R_j + p_j, R_i) + p_i - d_i, 0)]\} \end{aligned}$$

Therefore,

$$\begin{aligned} C_{ij} - C_{ji} = & \{[R_i + p_i - r_i] + [max(R_i + p_i - d_i, 0)]\} + \\ & \{[max(R_i + p_i, R_j) + p_j - r_j] + [max(max(R_i + p_i, R_j) + p_j - d_j, 0)]\} \\ & - \{[R_j + p_j - r_j] + [max(R_j + p_j - d_j, 0)]\} \\ & - \{[max(R_j + p_j, R_i) + p_i - r_i] + [max(max(R_j + p_j, R_i) + p_i - d_i, 0)]\} \\ = & \{[R_i + p_i - r_i] + [max(R_i + p_i, R_j) + p_j - r_j]\} \\ & + \{[max(R_i + p_i - d_i, 0)] + [max(max(R_i + p_i, R_j) + p_j - d_j, 0)]\} \\ & - \{[R_j + p_j - r_j] + [max(R_j + p_j, R_i) + p_i - r_i]\} \\ & - \{[max(R_j + p_j - d_j, 0)] + [max(max(R_j + p_j, R_i) + p_i - d_i, 0)]\} \end{aligned}$$

Now, with *PRTF* and $Q_1$ defined as previously, Chu [20] proved that $[R_i + p_i - r_i] + [\max(R_i + p_i, R_j) + p_j - r_j] = \max(PRTF(i,t), Q_1(i,j,t)) + (p_i + p_j) - (r_i + r_j).$

He also proved in [21] that with the above expression of *PRTT* and $Q_2$, the following equality is obtained

$[\max(R_i + p_i - d_i, 0)] + [\max(\max(R_i + p_i, R_j) + p_j - d_j, 0)] =$
$\max(Q_{ij}, PRTT(i,t) + (p_i + p_j)) - (d_i + d_j)$

where

$$\begin{aligned} Q_{ij} &= \max(R_i + p_i, d_i) + \max(R_j + p_j, d_j) \\ &= \max(R_i, d_i - p_i) + \max(R_j, d_j - p_j) + (p_i + p_j) \\ &= Q2(i,j,t) + (p_i + p_j). \end{aligned}$$

So,

$[\max(R_i + p_i - d_i, 0)] + [\max(\max(R_i + p_i, R_j) + p_j - d_j, 0)] =$
$\max(Q_2(i,j,t), PRTT(i,t)) + (p_i + p_j) - (d_i + d_j).$

By using these different equalities, it is obtained that

$$\begin{aligned} C_{ij} - C_{ji} &= \max(PRTF(i,t), Q_1(i,j,t)) + \max(Q_2(i,j,t), PRTT(i,t)) \\ &\quad - \{\max(PRTF(j,t), Q_1(j,i,t)) + \max(Q_2(j,i,t), PRTT(j,t))\} \\ &= FTR(i,j,t) - FTR(j,i,t). \end{aligned}$$

Hence task *i* dominates task *j* if and only if $FTR(i,j,t) \leq FTR(j,i,t)$ ∎

***Remark 4:*** The *FTR* function is a local optimality priority rule for the cost expressed as the sum of the total flow-time and the total tardiness penalty.

***Remark 5:*** It can be noted that if the cost is the weighted sum of the flow-time cost and the tardiness cost, where the flow-time is weighted by $W_f$ and the tardiness is weighted by $W_t$, then theorem 1 still holds if function *FTR* is replaced by the following priority function.

$FTRW(i,j,t) = W_f \cdot \max(PRTF(i,t), Q1(i,j,t)) + W_t \cdot \max(PRTT(i,t), Q2(i,j,t))$

Therefore, all the development that follows can still be used to solve the real-time maintenance tasks scheduling problem. These definitions and properties of the *FTR* function, combined with the dominance matrix Ω and the task strength, are then used in a scheduling algorithm that minimizes the objective function. This algorithm, designed for off-line implementation, is summarized as follow.

### 3.2. FTR-based algorithm

*Algorithm 1:* Off-line use of the *FTR* with one processor

*The decision instant t is the time when the processor is available the after the completion of a task k. The decision is made through the following steps.*

1) *The search set* $S_t \leftarrow \{all\ tasks\ not\ yet\ scheduled\}$.

2) *Compute FTR(i,j,t) for all pairs of tasks* $\{i, j\}$ *where* $i \in S_t$ *and* $j \in S_t$.

3) *Compute the dominance matrix* Ω *in the set* $S_t$.

4) *Compute the tasks' strength (the vector containing $f_i$ for all the tasks not yet scheduled) in the set* $S_t$.

5) *Evaluate the subset $S_{t_1}$ of tasks having the greatest strength.*

$S_t \leftarrow S_{t_1}$.

*Until card($S_t$)=1, go to step 4.*

*These steps are repeated until all tasks are scheduled.*

**Remark 6:** This algorithm can be adapted to the case of $q$ ($q \geq 2$) parallel processors. In this case, the processor available the earliest is used at each decision date.

**Proposition 3:** The algorithm based on the *FTR* function and the tasks' strength has an $O(n^3)$ complexity if $n$ is the number of tasks to be scheduled.

**Proof.** The proof of this proposition is straightforward.

A lower bound to the cost enables evaluating the performance of this algorithm and the priority rule on one processor. This lower bound is based on the SRPT rule (shortest remaining processing time rule) which gives a lower bound to the mean flow-time cost in presence of the unequal release dates of tasks and a lower bound to the mean tardiness cost in presence of the unequal release dates developed in [22] using modified due dates.

**3.3 Lower bound**

To obtain the lower bound, the non-preemtion assumption is relaxed and so interruption is permitted when the tasks are being processed. This is obtained in the following steps.

1) *Determine the series ($d'_1, d'_2,..., d'_n$) of the tasks' due dates sorted in their non decreasing order (modified due dates).*

2) *Schedule the preemptive tasks according to the SRPT rule. In the SRPT rule, tasks are scheduled in the non decreasing order of their remaining processing time.*

The quantity

$$\sum_{i=1}^{n} (c_{[i]i} - r_{[i]}) + \sum_{i=1}^{n} max(c_{[i]} - d'_i, 0)$$

is then a lower bound to the cost expressed as the sum of the flow-time and the tardiness. In this expression, [$i$] is the task completed in the $i^{th}$ position and $c_{[i]}$ its completion time. This lower bound is due to the fact that the SRPT rule gives a lower bound to the flow-time in presence of tasks with unequal release dates and, as proved in [22], the modified due dates combined with the SRPT rule gives a lower bound to the tardiness in presence of unequal release dates.

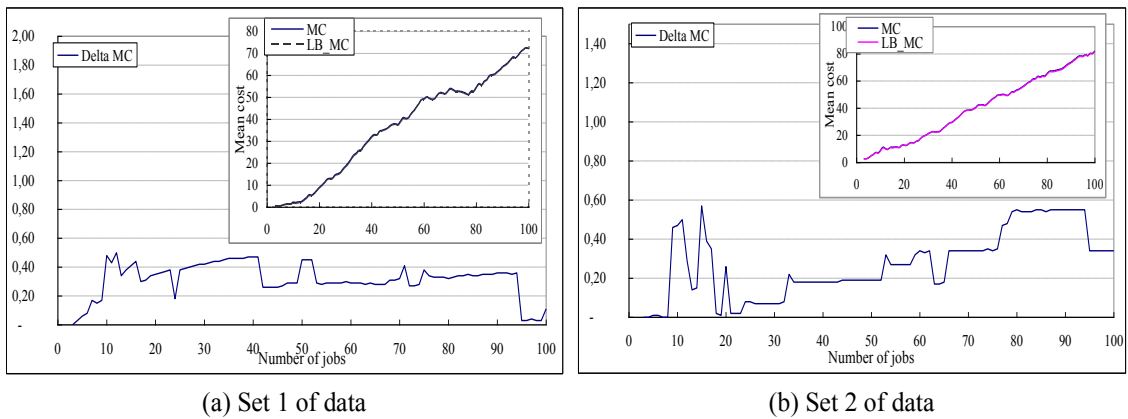

(a) Set 1 of data          (b) Set 2 of data

Fig. 1. The performance of the FTR rule relatively to the lower bound.

Before using this rule to solve the real-time maintenance problem, figure 1 shows some results obtained on two sets of data among others. On the curves, MC denotes the average cost (including the flow-time cost and the tardiness cost) obtained by using the *FTR* function and the dominance matrix. LB_MC is the lower bound to the average cost. Delta_MC is the difference (MC – LB_MC). These results, among others, reveal a very good performance of the rule.

Now, one considers the original problem which is the real-time preventive maintenance tasks scheduling and assignment to processors in order to minimize the total cost. It is proposed, in the following section, a solution based on the FTR function and the $\Omega$ matrix.

## 4. REAL-TIME APPLICATION AND EXPERIMENTATION

In addition to the development in the previous section, the properties below are used in the real-time decision-making process.

### 4.1 Additional properties

The following theorem allows the use of the *FTR* function in accordance with the precedence constraint linking the tasks on a machine.

***Theorem 2:*** *By using theorem 1, the precedence relationships linking the tasks on a given machine $E_{ik}$ are also respected. In other words, for any pair of integers $m_1$ and $m_2$ with $m_1 < m_2$, and for any machine $E_{ik}$, theorem 1 allows to schedule the $m_1^{th}$ task on $E_{ik}$ before the $m_2^{th}$.*

***Proof:*** The proof is based on the *FTR* function's definition.

Consider two tasks $i$ and $j$ whose processing times, release dates and due dates are such that $p_i = p_j$, $r_i \leq r_j$ and $d_i \leq d_j$. The *FTR* function defined in equation (9) is

$$FTR(i, j, t) = max[PRTF(i, t), Q_1(i, j, t)] + max[PRTT(i, t), Q_2(i, j, t)]$$

where

$PRTF(i, t) = 2.max(r_i, t) + p_i,$

$PRTT(i, t) = max(r_i, t) + max(max(r_i, t) + p_i, d_i),$

$Q_1(i, j, t) = max(r_i, t) + max(r_j, t),$ and

$Q_2(i, j, t) = max(max(r_i, t), d_i - p_i) + max(max(r_j, t), d_j - p_j).$

Then, $PRTF(i, t) \leq PRTF(j, t)$ and $PRTT(i, t) \leq PRTT(j, t)$.

As $Q_1(i, j, t) = Q_1(j, i, t)$ and $Q_2(i, j, t) = Q_2(j, i, t)$, then the inequality $FTR(i, j, t) \leq FTR(j, i, t)$ is satisfied. Now for two integers $m_1$ and $m_2$ with $m_1 < m_2$, the $m_1^{th}$ task and the $m_2^{th}$ task on a machine $E_{ik}$ are supposed to have the same processing time $Mtp_{ik}$ and they satisfy $r_{m_1,ik} \leq r_{m_2,ik}$ and $d_{m_1,ik} \leq d_{m_2,ik}$. So the *FTR* function gives the priority to the $m_1^{th}$ task.

The problem then becomes establishing priorities between the different tasks on machines in the system.

***Definition 4:*** Let $t$ be a decision-making time and $\Delta_{ik}(t)$ denote, at time $t$, the duration between $t$ and the end of the last preventive task on machine $E_{ik}$ before that time. Let $U_t$ (set of urgent tasks) denote at time $t$ the subset of tasks satisfying condition $\Delta_{ik}(t) \geq \tau_{ik}^1$ .

On the basis of these different definitions and properties, one proposes, below, a real-time algorithm for scheduling and assignment of tasks to processors in a distributed system. The algorithm uses the FTR function that takes into account the critical state costs and the tardiness costs, the dominance matrix $\Omega$, the strength vector, the urgency of tasks and theorem 2.

**4.2. Real-time decision making**

The real-time decision making consists in establishing priorities between the tasks, as they arrive in the system, and to assigning them to the processors. The algorithm, called OL-MTSA-2T, is summarized as follows.

*Algorithm 2:* OL-MTSA-2T

*At the beginning of the planning horizon H (t = 0):*

*Determine the first critical dates and the due dates of the preventive maintenance tasks on each machine in the system (they are equal to $\tau_{ik}^1$ and $\tau_{ik}^2$ respectively for machine $E_{ik}$).*

*The decisions are then made through the steps below.*

1) *Consider the time t when a processor is available the earliest.*

*If two or more processors are available at time t, select the one with the smallest index.*

*While t<H, go to step 2.*

*Else, go to step 5.*

2) *Compute the subset $U_t$.*

3) *If $card(U_t) \geq 1$ then the search set $S_t \leftarrow U_t$.*

*Else $S_t \leftarrow \{all\ the\ machine\ of\ all\ the\ sites\}$.*

4) *Apply to the set $S_t$ the steps described in algorithm 1 based on the FTR function for the tasks scheduling, to select the task that should be scheduled at time t.*

*Determine its completion time.*

*Determine the next release date of preventive maintenance task on the corresponding machine by adding the quantity $\tau_{ik}^1$ to its completion time.*

*Determine the next due date by adding its quantity $\tau_{ik}^2$ to its completion time.*

*Go to step 1.*

5) *End.*

The following subsection gives some experimental results highlighting the performances of the approach.

**4.3. Numerical experimentation**

The algorithms based on the *FTR* rule are programmed in C language and tested on a Compaq AlphaServer ES40 DEC6600 workstation operating under UNIX with 2048 MB of RAM memory. Two versions of the OL-MTSA-2T algorithms have been programmed. The first one uses the urgency criterion while the second one does not. Both programs are implemented on a 365-day horizon.

Experimentations are performed on systems with an overall number of 500 machines. The failure rates data ($\lambda_{ik}$) and the repair rates data ($\mu_{ik}$) are simulated according to normal laws and the processing time ($MTP_{ik}$) of tasks have been computed on the basis of the repair rates. The first availability thresholds $\alpha_{ik}^1$ are uniformly generated in the interval $]0, 1[$ with respect to the asymptotic availability $\mu_{ik}/(\lambda_{ik} + \mu_{ik})$. The durations $\tau_{ik}^1$ are then computed on the basis of these thresholds. Instead of the second thresholds ($\alpha_{ik}^2$), we generated the durations ($\tau_{ik}^2$) which are the duration from the end of a maintenance task on machine $E_{ik}$ to the deadline of the next task on the same machine. To obtain $\tau_{ik}^2$, we generated variables $\varepsilon_{ik}$ uniformly in [0.25, 0.5]. Then $\tau_{ik}^2$ is computed as $\tau_{ik}^2 = \tau_{ik}^1 + Mtp_{ik} + \varepsilon_{ik}.Mtp_{ik}$. The total number of processors is varied from 2% to 20% of the total number of machines in the system. It is noted that in the manufacturing industry, the number of processor is scarcely greater than $0.04 \times \sum_{k=1}^{N} N_k$. On graph (a) of figure 2, MC is the average cost of the preventive maintenance tasks undertaken on the 365-day horizon $H$. It corresponds to the average cost relating only to the machines whose tasks have been processed on the horizon $H$. MRC is the average cost of all the preventive tasks that have been needed on the horizon $H$ (including machines whose maintenance has not been able to be processed on the horizon). The _U (resp. _NU) extension to the curve names denotes the result obtained with (resp. without) considering the urgency criterion.

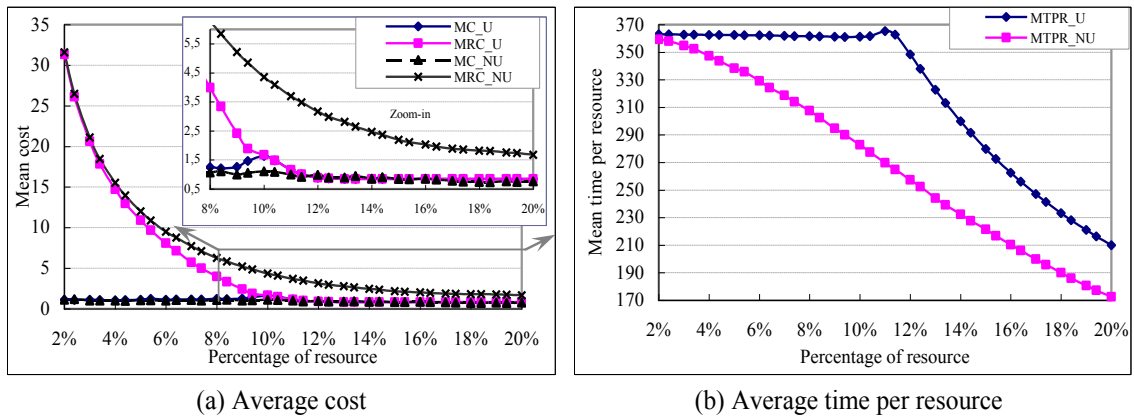

(a) Average cost  (b) Average time per resource

Fig. 2. Results on 500 machines

From this figure, one can observe the following. 1) The average cost of the tasks that are processed on the time horizon (MC) is almost the same whatever the tasks' urgency is used or not. But, on the overall set of tasks, including those that have not been able to be processed on the horizon, the average cost (MRC) is far better when the urgency is used. It represents the average cost involved for the whole system on the horizon including the cost of critical states and of tardiness. Moreover, on the enlarged part of the graph (a), one can see that when the number of resources or processors increases, the MC and the MRC (in the case where the urgency is used) converge from about 10%. This means that from a given number of resources, all the preventive tasks on the horizon are processed when the urgency is used (while minimizing their total cost, which is equivalent to minimizing the average cost). This is not the case when the urgency is not used.

2) Graph (b) of figure 2 shows the average time of processors utilisation on the 365-days time horizon. When the urgency criterion is considered, the processors are used at their full capacity as long as all the maintenance tasks are not processed. The average time per processor under the urgency criterion begins to decrease only from 10%, when all the tasks are processed. On the contrary, the mean time per resource continually decreases as the number of resources increases when the urgency is not used, even though there remain tasks that are not processed.

All these remarks emphasize the effectiveness of the priority rule expressed by the FTR function, when it is combined to the urgency criterion of tasks for making the preventive maintenance decisions in real time.

## 5. CONCLUSION

In this paper, we proposed a real-time algorithm for the scheduling of the preventive maintenance tasks of a distributed system comprising several sites, and each one containing a set of machines working in series. The main objective is to minimize the maintenance cost on a given horizon, while ensuring the system an availability higher than a given threshold. The maintenance cost comprises the cost of the time spent by the machines operating in critical states (expressed as the flow-time cost or response time cost), the cost due to the maintenance tasks tardiness and tasks processing cost. For this purpose, we propose a local optimality priority rule (FTR) using a priority rule for the flow-time minimization and a priority rule for the tardiness' minimization. This priority rule considers task pairs at decision times. This priority rule has a polynomial complexity of $O(n^3)$ when used for off-line scheduling. This polynomial complexity enables to use it as well in production scheduling as in real-time decision making of preventive maintenance tasks on large systems. We also proposed a lower bound which reveals the good performance of the rule with one processor. Basing on this rule and other concepts introduced in the paper, we proposed an algorithm for the real-time decision making in the preventive maintenance activities management of distributed systems.

Among the different assumptions, we considered exponential availability law for the machines incorporated in the system. This assumption is not restrictive. The method described in this paper can be used in the case of other laws. We also supposed that the processors are identical. In the extension to this work, non-identical processors will be considered.